\documentclass[runningheads]{llncs}
\usepackage{graphicx}

\usepackage{amsmath}
\usepackage{amssymb}
\usepackage{amsthm}
\usepackage{amstext}

\usepackage{booktabs}
\usepackage{tikz}
\usetikzlibrary{arrows, automata,calc}
\usepackage{graphicx}
\usetikzlibrary{decorations.text}
\usepackage{subcaption}
\usepackage{comment}
\usepackage{array}
\newcolumntype{P}[1]{>{\centering\arraybackslash}p{#1}}
\usetikzlibrary{shapes.arrows,calc,quotes,babel}
\usepackage{tikzpeople}
\usetikzlibrary{positioning}
\usepackage[export]{adjustbox}
\usepackage{pgfplots}
\pgfplotsset{compat=newest}
\tikzset{%
  Line Width/.code={%
    \pgfpointxy{#1}{0}%
    \pgfgetlastxy\tmpx\tmpy\tikzset{line width/.expanded=\tmpx}%
  },
  person/.style={
    Line Width=1/4, draw=black, color=gray!50
  },
  highlight/.style={
    preaction={Line Width=1/4, draw=gray!50, path fading=west, fading angle=-45},
    Line Width=1/4, draw=white, , path fading=east, fading angle=-45,
}}
\tikzset{>=latex}

\definecolor{USDT}{HTML}{EB5E55}
\definecolor{USDC}{HTML}{F6E27F}
\definecolor{BTC}{HTML}{E2C391}
\definecolor{ETH}{HTML}{A8B7AB}
\definecolor{DAI}{HTML}{9BBEC7}

\definecolor{blue1}{HTML}{072f5f}
\definecolor{blue2}{HTML}{1261a0}
\definecolor{blue3}{HTML}{3895d3}
\definecolor{blue4}{HTML}{58cced}

\usepackage{float}
\newcommand{\repeatthanks}{\textsuperscript{\thefootnote}}

\begin{document}
\title{An Empirical Study of Market Inefficiencies in Uniswap and SushiSwap}
%
%
\author{Jan Arvid Berg \and Robin Fritsch\thanks{Corresponding authors.} \and
Lioba Heimbach\repeatthanks \and Roger Wattenhofer}
\authorrunning{J. Berg et al.}
%
\institute{ETH Z{\"u}rich \\
\email{\{bergar,rfritsch,hlioba,wattenhofer\}@ethz.ch}}
%
%
\maketitle              
\begin{abstract}
    
    Decentralized exchanges are revolutionizing finance. With their ever-growing increase in popularity, a natural question that begs to be asked is: how efficient are these new markets?
    
    We find that nearly 30\% of analyzed trades are executed at an unfavorable rate. Additionally, we observe that, especially during the DeFi summer in 2020, price inaccuracies across the market plagued DEXes. Uniswap and SushiSwap, however, quickly adapt to their increased volumes. We see an increase in market efficiency with time during the observation period. Nonetheless, the DEXes still struggle to track the reference market when cryptocurrency prices are highly volatile. During such periods of high volatility, we observe the market becoming less efficient -- manifested by an increased prevalence in cyclic arbitrage opportunities.

\end{abstract}
\section{Introduction}
Nakamoto introduced the first fully decentralized cryptocurrency, Bitcoin~\cite{nakamoto2008bitcoin}, in 2008. In the following years, several blockchains followed, notably Ethereum~\cite{buterin2013ethereum} which introduced smart contracts. In their initial phase, blockchains only had a few niche applications, and the excitement surrounding them was mainly fueled by the hopes of continuously rising cryptocurrency prices.

However, this changed with the introduction of \emph{decentralized finance (DeFi)}. Suddenly, blockchains had a new purpose: offering financial services without the need of a middleman. \emph{Decentralized exchanges (DEXes)}, which allow users to trade in a fully noncustodial manner, are a main pillar of DeFi. Instead of requiring traders to give up custody of their funds by depositing into a centralized exchange (CEX), traders can now directly swap tokens with a smart contract on the blockchain. The popularity of DEXes is undeniable. 
The trading volume on all DEXes, which include Uniswap, SushiSwap, Balancer, Bancor, and Curve, exceeded \$50 billion in January 2021 and in every month since~\cite{defiprime}.

Most of these DEXes use a novel market-making mechanism. Rather than matching orders using a limit order book like traditional exchanges, most DEXes use an automated market maker mechanism that executes orders against a \emph{liquidity pool} holding token reserves.
The exchange rate in a liquidity pool is determined by a trading function and the amount of funds in the pool. Temporarily, these rates can be inaccurate and vary across different DEXes.
Such inaccuracies in the price lead to trades being executed at unfavorable rates if traders are not alert. Further, they can create cyclic arbitrage opportunities. Cyclic arbitrage opportunities indicate erroneous rates and thereby stem from lacking market efficiency, which measures how well the prices reflect all relevant information~\cite{malkiel1989stock}. As the market is becoming increasingly complex with an ever-growing number of DEXes and liquidity pools, studying the market's efficiency is ever more important. 

We investigate the existence and severity of market inefficiencies in two market leading DEXes, Uniswap and SushiSwap, between 12 September 2020 and 23 January 2021. With the optimal routing problem, we find trades that executed with an unfavorable rate -- indicating both the presence of price inaccuracies in the market and their effects on traders. Further, we look for past cyclic arbitrage opportunities, which stem from price differences in the market, and use them as a tool to determine how efficient the market as a whole is.

\section{Related Work}

Numerous studies analyze market efficiency on CEXes~\cite{stiglitz1982inefficiency,malkiel1989stock,bernard1997accounting,zunino2008multifractal}. Conversely, we measure the market efficiency on DEXes, by studying the prevalence of price inaccuracies across the market. More precisely, we focus on DEXes using an \emph{automated market maker (AMM)}.

AMMs have been around for quite some time, e.g., the logarithmic market scoring rule used in prediction markets~\cite{hanson03}. However, only the recent emergence of DEXes has made AMMs a popular alternative to traditional central limit order book systems. Uniswap and SushiSwap, along with most other DEXes, implement a new type of AMM design called \emph{constant function market maker (CFMM)}~\cite{angeris20oracles}. More specifically, Uniswap and SushiSwap use a form of CFMM called \emph{constant product market maker} (CPMM). Angeris et al.~\cite{angeris2019analysis} formally analyze how closely CPMMs track reference markets and demonstrate the numerical stability of CPMMs under a wide range of market conditions. We, on the other hand, empirically study the efficiency of Uniswap and SushiSwap in tracking the reference market. By looking for optimizable trades and cyclic arbitrage opportunities, both caused by price inaccuracies in the market, we identify moments when some prices on the DEXes do not accurately track the reference markets. 

Several works quantify and study \emph{Blockchain Extractable Value (BEV)} in DeFi. BEV measures the extractable profit from the blockchain by including, excluding, and re-ordering the transactions in a block. A broad study on both the amount of extractable and extracted BEV in DeFi is provided by Qin et al.~\cite{qin2021quantifying}. Daian et al.~\cite{daian2020flash} study front-running in decentralized exchanges and show that these arbitrage opportunities can cause a priority gas auction, which drives up the gas fee for all blockchain users. Through studying back-running on DEXes, Zhou et al.~\cite{zhou2021a2mm} show that this kind of BEV can cause back-run flooding, a denial of service practice on the blockchain. As opposed to focusing on the effects of arbitrage opportunities on DEXes, we empirically study one of their root causes -- price inaccuracies in the market. 

Danos et al.~\cite{danos20} theoretically study the optimal routing of trades through a network of CFMMs. We apply the special case of independent paths studied by Danos et al.~\cite{danos21} to analyze past trading data. In examining big swaps, we find that a significant fraction executes suboptimally on Uniswap and SushiSwap. Traders are, thus, suffering from price inaccuracies across the market. 

Wang et al.~\cite{wang2021cyclic} focus on cyclic arbitrage trades, a type of BEV, analyzing them theoretically and empirically. In contrast to this work, we study the availability of cyclic arbitrage opportunities in this paper and use it to identify price inaccuracies in the market. We further look for the causes of the inaccuracies and use them as a tool to study the market's efficiency over time.

\section{Constant Product Market Makers}

CPMMs, which include Uniswap and SushiSwap, are smart contracts running on the Ethereum blockchain.
On a CPMM, anyone can create a liquidity pool for an arbitrary token pair. Once created, liquidity providers deposit amounts of equal value of both tokens into the pool.
Traders swap tokens with the pool's liquidity and pay a small trading fee, which is distributed pro-rata among all liquidity providers.
To execute such a transaction on the blockchain, traders first submit their transaction to the Ethereum \emph{mempool} and wait for a miner to include them in the next block.

The CPMM smart contract determines the output amount the trader receives. The amount returned by the CPMM ensures that the product of the two reserve amounts stays constant. More precisely, consider a pool that holds reserves of a token $A$ and token $B$. Assume the reserves are $R_A$ token $A$ and $R_B$ token $B$, and that a trader wants to swap $t_A$ token $A$. The trader then receives 
\begin{equation*}
    t_B= R_B - \frac{R_AR_B}{R_A+(1-f)t_A} = \frac{R_B(1-f)t_A}{R_A+(1-f)t_A},
\end{equation*}
where $f$ is the transaction fee charged on the input amount (0.3\% in Uniswap)~\cite{adams2020uniswap}. Thus, when swapping an amount of $t_A$ token $A$, the price per token $B$ is
$\frac{R_A+(1-f)t_A}{R_B(1-f)}$.
We note that the price per token is increasing in the input amount. The larger the trade gets, the more the trader has to pay per desired token. This effect is a consequence of a trade's \emph{price impact} -- the impact of an individual trade on the market price. We also see in the formula above, that a trade's price impact is lower in pools with larger reserves. Thus, it is both the ratio of its assets and the liquidity of the pool that determine the price. 


\section{Data Description}
We analyze data from Uniswap and SushiSwap: Ethereum's two largest DEXes by trading volume. Together they represent the majority of the market, as they account for more than half of Ethereum's DEX trading volume~\cite{defiprime}. Further, both use the exact same trading mechanism, making them ideal for analyzing price inaccuracies between markets. More precisely, we collect pool reserve and transaction data between 5 August 2020 and 23 January 2021 for all pools between any of the following five cryptocurrencies: Bitcoin (BTC), Ethereum (ETH), Tether (USDT), USD Coin (USDC), and Dai (DAI). These are some of the most traded tokens and are among those which share the highest number of pools with other cryptocurrencies~\cite{Heimbach2021behavior}. Thus, they are well suited for our analysis. The high number of pairwise pools provides a large set of independent paths between cryptocurrency pairs to find optimizable routes and cyclic arbitrage opportunities.
For a thorough data description, see Appendix~\ref{app:data}.

\section{Identifying Market Inefficiencies}
To find market inefficiencies and study how they evolved over time, we analyze both the optimizability of past transactions and the opportunity for cyclic arbitrage. Price differences, the root cause of suboptimal trades, and cyclic arbitrage opportunities reflect the market's ability to reflect the relevant information. 
\subsection{Suboptimal Trade Routing}\label{sec:traderouting}

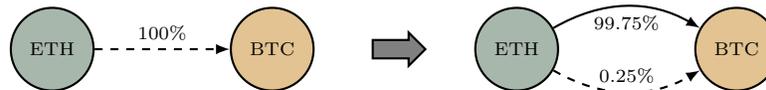
\begin{figure}[htbp]\vspace{-6mm}
\centering
    \begin{tikzpicture}[scale = 0.65]

    \tikzstyle{vertex1} = [circle, fill=blue!10]
    \tikzstyle{vertex2} = [circle, fill=red!10]
    \tikzstyle{vertex3} = [circle, fill=green!10]
    \tikzstyle{edge1} = [->, thick]
    
    \node[draw, thick,circle, minimum size = 1.1cm, fill =ETH](v1) at (-4.5,0){\scriptsize ETH};
    \node[draw, thick,circle, minimum size =  1.1cm, fill =BTC](v2) at (0,0){\scriptsize BTC};
    
    \node[draw, thick,circle, minimum size =  1.1cm, fill =ETH](v3) at (5,0){\scriptsize ETH};
    \node[draw, thick,circle, minimum size =  1.1cm, fill =BTC](v4) at (9.5,0){\scriptsize BTC};
    \node at (2.5,0) [thick, fill = gray,
        draw=black, 
        single arrow,
        minimum width=0.3cm, 
        minimum height=0.7cm, 
        single arrow head extend=1mm,
        rotate=0
    ] {};
    
    \path[->, draw]
    (v1) edge[dashed, thick] node[sloped, anchor=center, align = center, above, text width=2.0cm]  {\scriptsize 100\%}(v2)
    
    (v3) edge[ bend left = 30, thick] node[sloped, anchor=center, align = center, below, text width=2.0cm] {\scriptsize 99.75\%}(v4)
    (v3) edge[dashed, bend right = 30, thick] node[sloped, anchor=center, align = center, above, text width=2.0cm] {\scriptsize 0.25\%}(v4)
    ;
    \end{tikzpicture}
\caption{Example of original and optimizable swap. The Uniswap pools are represented by solid lines and SushiSwap pools are represented by dashed lines.}\label{fig:tradeopt}\vspace{-6mm}
\end{figure}

We start with an example of a trade that was executed at an unfavorable rate to illustrate the meaning of suboptimal trade routing (Figure~\ref{fig:tradeopt}). On 15 November 2020, the original trade exchanged ETH for 16.61 BTC. The trade routed through the ETH-BTC SushiSwap pool, as we show on the left in Figure~\ref{fig:tradeopt}. We visualize an optimization that executes 99.75\% of the trade in the Uniswap pool and only 0.25\% of the trade in the SushiSwap pool. By routing the transaction as shown, the trader would have received 18.07 BTC -- an 8.79\% improvement. We notice that the Uniswap trade offered a better price for this trade. Thus, the market was unsynced at this moment, and thereby prices in at least one pool did not reflect all relevant information -- indicative of market inefficiency. As a result of the large trade size, it was optimal to route a small part of the trade through the SushiSwap pool to reduce the trade's price impact. 

We analyze past transactions over \$30'000 from 12 September 2020 to 23 January 2021 and check whether they were routed optimally.
We focus on large trades since they have a more significant price impact, and the benefit from an optimal routing is greater for these trades compared to smaller ones. These trades are, thus, particularly well suited to study in order to detect inefficiencies in the market.
We also note that we exclude swaps that are part of a transaction including multiple swaps. While some of these swaps could be optimizable, others might already be part of optimized routing. Then optimizing one of them without considering the others would overestimate the possibility for optimization.

To identify trades that received unfavorable prices, we apply a special case of the optimal trade routing problem studied by Danos et al.~\cite{danos21}. They provide the solution to the optimal routing problem for trades through a set of independent paths. The solution accounts for the CPMM's transactions fees. We construct the set of independent paths as follows: we include both direct routes (Uniswap and SushiSwap) if they exist. The direct route is the pool containing the input and output tokens. For paths containing multiple pairs, we consider the more liquid Uniswap pool (Figure~\ref{fig:liq_pools}). We search for transactions whose output can be increased by more than \$30 with optimized routing and consider them optimizable. The threshold accounts for additional gas fees related to potentially routing the trade through several routes. The potential for such optimization indicates either the existence of price inaccuracies or that the market is illiquid. 

29'611 out of 108'667 analyzed transactions (Figure~\ref{fig:pools}) were optimizable --  a share of 27\%. On average, it is possible to increase the output of an optimizable trade by 0.15\% (Table \ref{tab:average_gain}) by using an optimized routing. While it might not appear to be a significant proportion, the 0.15\% is an invisible tax placed on trades stemming from price inaccuracies present in the market. We further observe that the mean gain is considerably higher than the median gain, suggesting that the tax on the most affected traders is significantly higher. Looking at the top 5\% trades, where optimized routing would improve by price most significantly (in percentage terms), we find that the average achievable gain was 0.71\% (Table \ref{tab:average_gain}). Further, indicating that traders are suffering from these market inefficiencies.

\begin{figure}[htbp]\vspace{-6mm}
    \centering
    \begin{tikzpicture}[scale = 0.65]
		\draw[fill=ETH] (0,0) rectangle (3.174,0.8) node[pos=.5] {$1$};
		\draw[fill=USDT] (3.174,0) rectangle (13.05,0.8) node[pos=.5] {$2$};
        \draw[fill=DAI] (13.05,0) rectangle (16.47,0.8) node[pos=.5] {$3$};
        \draw[fill=USDC] (16.47,0) rectangle (17.4,0.8) node[pos=.5] {$\geq4$};
	\end{tikzpicture}
    \caption{Proportions of number of paths used by optimized routings.}
    \label{fig:numberpaths}\vspace{-6mm}
\end{figure}
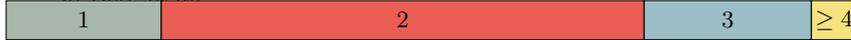

\begin{table}[htbp]\vspace{-6mm}
    \centering
    \footnotesize 
    \begin{tabular}{@{}p{2.2cm}   P{3cm} P{3.5cm}  @{}}
	    \toprule
		\textbf{} & \textbf{All Trades} &\textbf{Top 5\% Trades}\\
		\midrule
        Mean Gain& 0.15\% &  0.71\%  \\
        \bottomrule
	\end{tabular}\vspace{1mm}
    \caption{Gains achieved by optimized routings.}\label{tab:average_gain}\vspace{-10mm}
\end{table}
To check whether the market inefficiencies stem from price inaccuracies or the potential lack of liquidity in the market, we analyze how many paths were used by the optimized routings (Figure \ref{fig:numberpaths}). We count a path if at least 0.1\% of the trade routes through it. 18\% of optimizable trades only use a single path with the new routing. Thus, this optimized routing does not include the original path at all. The unfavorable price received, thereby, solely stems from price inaccuracies. We note that the optimized routing for a small proportion of trades consists of at least three paths. There are at most five possible paths in the network (Appendix~\ref{app:networks}, Figure~\ref{fig:liq_pools}). These relatively complex routings indicate that the liquidity of the pools also causes unfavorable rates received by large trades.

When we repeat the analysis and include the less liquid SushiSwap pools (Appendix~\ref{app:networks}, Figure~\ref{fig:less_liq_pools}) in the set of independent paths, we find that the optimization potential increases. We perform this adjacent analysis on a smaller set of 43'321 swaps, which include all trades originally executed in the following pools: USDC-ETH (Uniswap and SushiSwap) and DAI-ETH (SushiSwap). Here, we find that an even larger proportion of swaps, 33.75\% compared to 27\% previously, could have been optimized. Further, the average gains are higher in the less liquid pools (Table \ref{tab:6.6}). We find that routing through lower liquidity pools leads to better optimizing possibilities. Thus, it seems to be price inaccuracies across the market driving the unfavorable rates as opposed to the lack of liquidity in parts of the market -- indicative of the presence of market inefficiencies in (parts of) the market. The significant number of trades with unfavorable rates shows that traders are experiencing the consequences of these inefficiencies.



\begin{table}[htbp]\vspace{-6mm}
    \centering
    \footnotesize 
    \begin{tabular}{@{}p{2.2cm}   P{3.5cm} P{3.5cm}  @{}}
	    \toprule
		\textbf{}&\textbf{More Liquid Pools} & \textbf{Less Liquid Pools} \\
		\midrule
        Mean Gain  &0.15\% & 0.16\%   \\ 
        \bottomrule
	\end{tabular}\vspace{1mm}
    \caption{Gains achieved by optimized routings percent for the two sets of pools.}
    \label{tab:6.6}
\vspace{-12mm}\end{table}

\subsection{Cyclic Arbitrage Opportunities}

Cyclic arbitrage opportunities result from temporary price inaccuracies. By taking advantage of non-equilibrium exchange rates within and between markets, arbitrage traders can profit by trading their funds in a cycle. However, the existence of such opportunities suggests market inefficiency. The exchange rate of at least one pool in such a cycle must not accurately reflect all relevant information. Studying the prevalence and duration of these cycles brings inside new market insights beyond the unfavorable prices experienced by traders.

We identify cyclic arbitrage opportunities retrospectively by searching for directed cycles of pools where a cyclic swap with $\alpha$ tokens $A$, the returns $\hat{\alpha}$ tokens $A$ where $\hat{\alpha} > \alpha$. In a cycle  $c = (e_1,\dots,e_n)$ each edge $e_i$ present in the cycle represents a pool used in the transactions. For each arbitrage opportunity, we compute the maximum possible profit. The problem's convexity~\cite{danos20} lets us find the optimal, unique solution that maximizes the profit $\hat{\alpha} - \alpha$. 

\begin{figure}[htb!]\vspace{-5mm}
\centering
 \includegraphics[width=0.95\columnwidth,right]{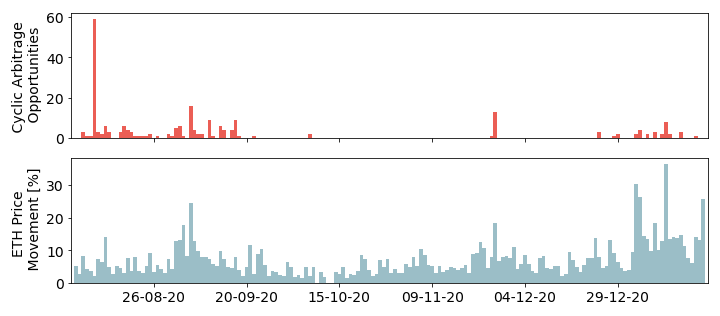}
\caption{Daily number of blocks with cyclic arbitrage opportunities and ETH price movement between 5 August 2020 and 23 January 2021.}%
\label{fig:arb_per_100k}%
\vspace{-6mm}
\end{figure}

Over six months, from 5 August 2020 to 23 January 2021, we analyze the occurrence of cyclic arbitrage opportunities between ETH, USDC, and USDT on Uniswap. These three cryptocurrencies have the largest pairwise Uniswap pools between them.
We plot the daily number of blocks with arbitrage opportunities exceeding 30\$ and ETH's daily price movement in Figure~\ref{fig:arb_per_100k}. Daily price movement is a volatility measure and given by is $100\% \cdot \frac{p_{\text{high}}-p_{\text{low}}}{p_{\text{low}}}$. Here, $p_{\text{high}}$ is the day's highest price and $p_{\text{low}}$ is the day's lowest price.

Most arbitrage opportunities are between August and September 2020 (Figure~\ref{fig:arb_per_100k}). In this period, Uniswap experienced an incredible increase in daily trading volume. Within six weeks, the volume increased steadily by around 700\%, reaching \$500 million at the end of September. Similar daily trading volumes were not consistently reached again until early January 2021. Arbitrage opportunities also experience a sharp decrease starting at the end of September, and we only begin to see a consistent uptake at the end of December. In general, the number of arbitrage opportunities appears to decrease throughout the analyzed period -- indicating that the market is becoming more efficient. We observe a second trend when comparing the number of daily arbitrage opportunities to ETH's price movement. With 0.15, the correlation between the two is relatively low. However, after the initial explosion in volume in the summer/autumn of 2020, we find that on days with exceptionally high price movements, we also observe exceptionally many arbitrage opportunities. 

To further investigate the influence of external cryptocurrency prices on the number of arbitrage opportunities, we search for cycles in five pools on Uniswap and SushiSwap between ETH, USDC, USDT, DAI, and BTC during the following two periods: (1) 11 September to 3 October 2020, and (2) 23 December 2020 to 23 January 2021. The first period analyzed is characterized by relatively stable cryptocurrency prices -- the price of ETH moved by less than 8\%. The second period, however, is characterized by highly volatile cryptocurrency prices -- the price of ETH more than doubled. By then, the pools containing BTC had accumulated enough liquidity, and we included them in the analysis. The analyzed networks during both periods are shown in Figures~\ref{fig:arb_cyc_4_coins} and \ref{fig:arb_cyc_5_coins} (Appendix~\ref{app:networks}).


\begin{table}[htbp]\vspace{-6mm}
    \centering
    \footnotesize 
    \begin{tabular}{@{}p{3cm}   P{4.1cm} P{2.3cm}  P{2.5cm}  @{}}
	    \toprule
		\textbf{} & \textbf{Blocks with Cyclic Arb.} &\textbf{Mean Profit}&\textbf{Avg. Duration}\\
		\midrule
        \textbf{11.09.20 - 03.10.20} &84 & 0.24\%  & 2.33 blocks  \\
        \textbf{23.12.20 - 23.01.21} & 1'061& 0.35\% &  1.43 blocks\\
        \bottomrule
        
	\end{tabular}\vspace{1mm}
    \caption{Statistics of cyclic arbitrage profits in the two analyzed periods.}
    \label{tab:stat_profit}
\vspace{-10mm}\end{table}

We find that 84 out of 140,000 blocks have arbitrage cycles with a mean profit of 0.24\% in the first period (Table~\ref{tab:stat_profit}). Stable cryptocurrency prices appear to not allow for many arbitrage opportunities, even in the face of the recent introduction of SushiSwap at the time. The markets' exchange rates are synced. In the second period, we find 1061 blocks with cyclic arbitrage opportunities, and, on average, they present a profit of 0.351\% (Table~\ref{tab:stat_profit}).

\begin{figure}[tbp]\vspace{-5mm}
   \centering
     \includegraphics[width=0.95\columnwidth,right]{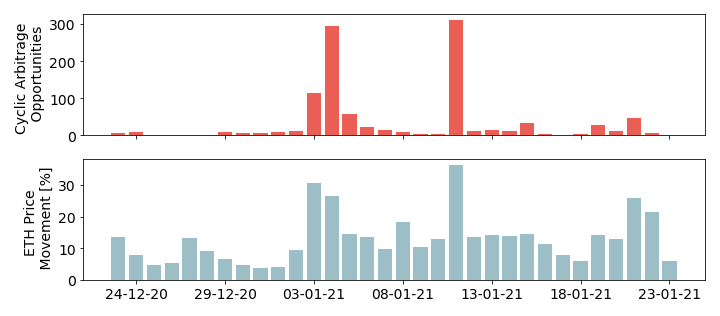}
  \caption{Daily number of blocks with cyclic arbitrage opportunities and ETH price movement from 23 December 2020 to 23 January 2021.}\label{fig:arb_2_period}
  \vspace{-6mm}
\end{figure}

As suspected, the number of cyclic arbitrage opportunities is significantly higher in the second period, where cryptocurrencies prices are highly volatile. To show the correlation, we plot the distribution of arbitrage opportunities and ETH's price movement in Figure \ref{fig:arb_2_period}. Two days, 4 January and 11 January 2021, make up for over 50\% of all identified opportunities. The price of ETH experienced temporary movements exceeding 20\% on both days. Further, the correlation between the number of arbitrage opportunities and the ETH price movement is 0.72. We conclude that, as expected, higher price volatility is more favorable for arbitrage. The floating exchange rates on Uniswap and SushiSwap do not adjust to market price sufficiently fast. For every cyclic arbitrage opportunity, there is at least one pool whose price does not reflect all relevant information. We also observe that arbitrage bots, who profit from the BEV caused by market inefficiencies, become more efficient in the months between the two periods. The arbitrage opportunities are available for 2.33 blocks on average in the first period and only available for 1.43 blocks on average in the second period (Table~\ref{tab:stat_profit}). Since the minimum availability of an arbitrage opportunity we observe is 1, this improvement in the efficiency of arbitrage bots is significant.\vspace{-2mm}

\section{Conclusion}
Traders are actively suffering from market inefficiencies on DEXes: nearly 30\% of analyzed trades were executed at an unfavorable price. We show that these erroneous rates stem largely from price inaccuracies across the market's liquidity pools. The market struggled to accurately reflect all relevant price information, especially during the initial volume explosion on DEXes in the late summer of 2020 -- evident from the increased number of cyclic arbitrage opportunities. However, the market quickly adapts, and the price inaccuracies largely disappear. Later, we only observe market inefficiencies on days when cryptocurrency prices are highly volatile. Parts of the market do not adjust to external price changes properly, thereby creating cyclic arbitrage opportunities. Arbitrage bots are also becoming more efficient and take advantage of these prevailing market inefficiencies within a block at most times. 

\newpage
%
%
%
\bibliographystyle{splncs04}
%
\bibliography{references}
\newpage
\appendix
\section{Data Description}\label{app:data}

In this section we provide a detailed data description. 

\subsection{Reserve Data}
By launching a go-ethereum client, we collect all pool reserves recorded on Ethereum from block 10000835 (4 May 2020, day of Uniswap V2 deployment) to block 11709847 (23 January 2021) for Uniswap. For SushiSwap, the client exports all pool reserves from block 10750000 (28 August 2020, day of SushiSwap deployment) to block 11709847 (23 January 2021).

\subsection{Transaction Data}
We collect the transaction data required for the optimal trading analysis from The Graph's Uniswap V2 subgraph\footnote{https://thegraph.com/explorer/subgraph/uniswap/uniswap-v2} and The Graph's SushiSwap subgraph\footnote{https://thegraph.com/explorer/subgraph/benesjan/sushi-swap} with GraphQL. The Graph is a decentralized protocol for indexing and querying data from blockchains like Ethereum. 

For each transaction between block 10000835 and block 11709847, we collect the amounts of tokens swapped, as well as the value of the swap in \$.
While the subgraphs we use also provide pool reserves, these are updated infrequently and therefore could not be used. For this reason, we collect pool reserves with the go-ethereum client.

We pre-process the transaction data to reflect the reserves at the beginning of each block. These reserves coincide with those at the end of the previous block and are the only information most traders use when submitting their transaction (assuming their transaction is included in the next block). Depending on other transactions executed ahead of the swap in the same block, the pool reserves might shift slightly before trade execution and thus change the exchange rate. We correct this effect by pre-processing the data accordingly such that the transactions reflect the trader's view -- based on which we optimize in Section~\ref{sec:traderouting}.

\subsection{Pool Network}

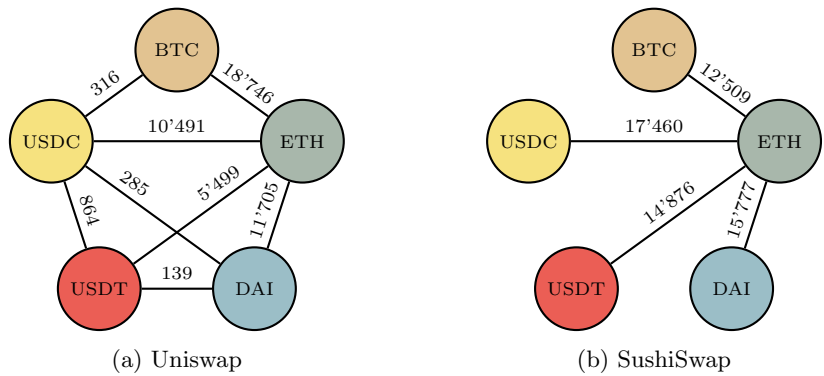
\begin{figure}[htbp]
  \begin{subfigure}{0.48\linewidth}
   \centering
    \begin{tikzpicture}[scale=0.65]
			\def \n {5}
			\def \radius {2.7cm}
			\def \margin {15} 

			\node[draw, thick,circle, minimum size = 1.1cm, fill =DAI] at ({360/\n *3+90}:\radius) (s1)  {\scriptsize DAI};
			\node[draw, thick,circle, minimum size = 1.1cm, fill =USDT] at ({360/\n *2+90}:\radius) (s2) {\scriptsize USDT};
			\node[draw, thick,circle, minimum size = 1.1cm, fill =USDC] at ({360/\n *1+90}:\radius) (s3) {\scriptsize USDC};
			\node[draw, thick,circle, minimum size = 1.1cm, fill =BTC] at ({360/\n *0+90}:\radius) (s4) {\scriptsize BTC};
			\node[draw, thick,circle, minimum size = 1.1cm, fill =ETH] at ({360/\n *4+90}:\radius) (s5) {\scriptsize ETH};

		    \path[thick]   (s3) edge	node[sloped, anchor=center, align = center, above, text width=2.0cm] 	{\scriptsize 10'491}	(s5)
                            (s2) edge  	node[sloped,  pos=.7,anchor=center, align = center, above, text width=2.0cm]   	{\scriptsize 5'499} 	(s5)
                            (s1) edge  	node[sloped, pos=.7,anchor=center, align = center, above, text width=2.0cm]   	{\scriptsize 285} 	(s3)
                            (s1) edge	node[sloped, anchor=center, align = center, above, text width=2.0cm] 	{\scriptsize 139}	(s2)
                            (s2) edge	node[sloped, anchor=center, align = center, above, text width=2.0cm] 	{\scriptsize 864}	(s3)
                            (s3) edge	node[sloped, anchor=center, align = center, above, text width=2.0cm] 	{\scriptsize 316}	(s4)
                            (s4) edge	node[sloped, anchor=center, align = center, above, text width=2.0cm] 	{\scriptsize 18'746}	(s5)
                            (s5) edge	node[sloped, anchor=center, align = center, above, text width=2.0cm] 	{\scriptsize 11'705}	(s1);

	\end{tikzpicture}
    \caption{Uniswap} \label{fig:pool_uni}
  \end{subfigure}%
\hfill
  \begin{subfigure}{0.48\linewidth}
    \centering
   \begin{tikzpicture}[scale = 0.65]
			\def \n {5}
			\def \radius {2.7cm}
			\def \margin {15} 

			\node[draw, thick,circle, minimum size = 1.1cm, fill =DAI] at ({360/\n *3+90}:\radius) (s1)  {\scriptsize DAI};
			\node[draw, thick,circle, minimum size = 1.1cm, fill =USDT] at ({360/\n *2+90}:\radius) (s2) {\scriptsize USDT};
			\node[draw, thick,circle, minimum size =1.1cm, fill =USDC] at ({360/\n *1+90}:\radius) (s3) {\scriptsize USDC};
			\node[draw, thick,circle, minimum size =1.1cm, fill =BTC] at ({360/\n *0+90}:\radius) (s4) {\scriptsize BTC};
			\node[draw, thick,circle, minimum size = 1.1cm, fill =ETH] at ({360/\n *4+90}:\radius) (s5) {\scriptsize ETH};

		    \path[thick]    (s3) edge	node[sloped, anchor=center, align = center, above, text width=2.0cm] 	{\scriptsize 17'460}	(s5)
                            (s2) edge  	node[sloped, anchor=center, align = center, above, text width=2.0cm]   	{\scriptsize 14'876} 	(s5)
                            (s1) edge	node[sloped, anchor=center, align = center, above, text width=2.0cm] 	{\scriptsize 15'777}	(s5)
                            (s4) edge  	node[sloped, anchor=center, align = center, above, text width=2.0cm]   	{\scriptsize 12'509} 	(s5);

	\end{tikzpicture}
    \caption{SushiSwap} \label{fig:pool_sushi}
  \end{subfigure}%

    \caption{The number of independent trades exceeding \$30'000 in each existing pool between any pair of the following cryptocurrencies: BTC, ETH, USDC, USDT and DAI, in Uniswap (Figure~\ref{fig:pool_uni}) and Suhiswap (Figure~\ref{fig:pool_sushi}). Data was collected between 12 September 2020 and 23 January 2021.}
    \label{fig:pools}
\end{figure}

Figure~\ref{fig:pools} visualizes the number of independent swaps exceeding \$30'000 in both Uniswap and Suhiswap between 12 September 2020 and 23 January 2021. In Uniswap we have a pool between almost every pair of tokens, while for Suhiswap the liquid pools during the observed time period all included ETH. Nonetheless, even on Uniswap most large trades are executed in pools that include ETH. 

\clearpage
\section{Networks}\label{app:networks}

We show the networks used to find optimized routing in Figure~\ref{fig:6.3} and the networks used to find cyclic arbitrage opportunities in Figure~\ref{fig:arb_cyc_coins}.
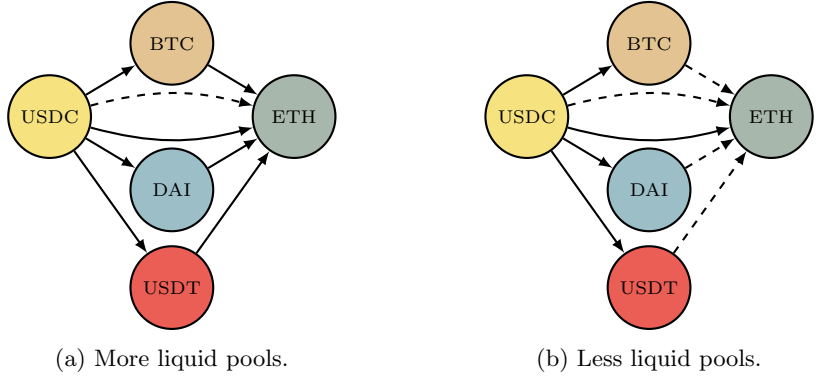
\begin{figure}[htbp]
  \centering
  \begin{subfigure}[t]{0.48\linewidth}
  \centering
     \begin{tikzpicture}[scale = 0.65]

\node[draw, thick,circle, minimum size =1.1cm, fill =USDC](v1) at (-2.5,0){\scriptsize USDC};
\node[draw, thick,circle, minimum size = 1.1cm, fill =BTC](v2) at (0,1.5){\scriptsize BTC};

\node[draw, thick,circle, minimum size = 1.1cm, fill =DAI](v3) at (0,-1.5){\scriptsize DAI};
\node[draw, thick,circle, minimum size = 1.1cm, fill =USDT](v4) at (0,-3.5){\scriptsize USDT};
\node[draw, thick,circle, minimum size = 1.1cm, fill =ETH](v5) at (2.5,0){\scriptsize ETH};


\path[->, draw]

(v1) edge[thick] node[sloped, anchor=center, align = center, above, text width=2.0cm] {}(v2)
(v1) edge[ thick] node[sloped, anchor=center, align = center, above, text width=2.0cm]  {}(v3)
(v1) edge[thick] node[sloped, anchor=center, align = center, above, text width=2.0cm]  {}(v4)

(v1) edge[thick, bend right=15] node[sloped, anchor=center, align = center, above, text width=2.0cm]  {}(v5)
(v1) edge[ thick, dashed, bend left=15] node[sloped, anchor=center, align = center, above, text width=2.0cm] {}(v5)

(v2) edge[thick] node[sloped, anchor=center, align = center, above, text width=2.0cm] {}(v5)
(v3) edge[ thick] node[sloped, anchor=center, align = center, above, text width=2.0cm]  {}(v5)
(v4) edge[thick] node[sloped, anchor=center, align = center, above, text width=2.0cm]  {}(v5);

\end{tikzpicture}
\caption{More liquid pools.}\label{fig:liq_pools}
  \end{subfigure}
  \hfill
    \begin{subfigure}[t]{0.48\linewidth}
    \centering
     \begin{tikzpicture}[scale = 0.65]

\node[draw, thick,circle, minimum size = 1.1cm, fill =USDC](v1) at (-2.5,0){\scriptsize USDC};
\node[draw, thick,circle, minimum size = 1.1cm, fill =BTC](v2) at (0,1.5){\scriptsize BTC};

\node[draw, thick,circle, minimum size = 1.1cm, fill =DAI](v3) at (0,-1.5){\scriptsize DAI};
\node[draw, thick,circle, minimum size = 1.1cm, fill =USDT](v4) at (0,-3.5){\scriptsize USDT};
\node[draw, thick,circle, minimum size = 1.1cm, fill =ETH](v5) at (2.5,0){\scriptsize ETH};

\path[->, draw]

(v1) edge[thick] node[sloped, anchor=center, align = center, above, text width=2.0cm] {}(v2)
(v1) edge[ thick] node[sloped, anchor=center, align = center, above, text width=2.0cm]  {}(v3)
(v1) edge[thick] node[sloped, anchor=center, align = center, above, text width=2.0cm]  {}(v4)

(v1) edge[thick, bend right=15] node[sloped, anchor=center, align = center, above, text width=2.0cm]  {}(v5)
(v1) edge[ thick,dashed, bend left=15] node[sloped, anchor=center, align = center, above, text width=2.0cm] {}(v5)

(v2) edge[dashed, thick] node[sloped, anchor=center, align = center, above, text width=2.0cm] {}(v5)
(v3) edge[dashed,  thick] node[sloped, anchor=center, align = center, above, text width=2.0cm]  {}(v5)
(v4) edge[dashed, thick] node[sloped, anchor=center, align = center, above, text width=2.0cm]  {}(v5);


\end{tikzpicture}
\caption{Less liquid pools.}\label{fig:less_liq_pools}
  \end{subfigure}
  
  \caption{Paths from USDC to ETH for both cases. Uniswap pools are represented by solid lines and SushiSwap pools are represented by dashed lines.}%
  \label{fig:6.3}%
\end{figure}
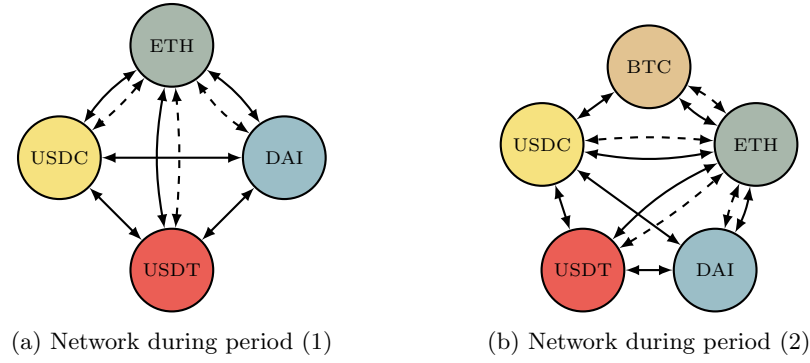
\begin{figure}[htbp]
    \centering
    \begin{subfigure}[b]{0.48\linewidth}
    \centering
    \begin{tikzpicture}[scale = 0.65]

    \def \n {4}
	\def \radius {2.3cm}
	\def \margin {15} 

	\node[draw, thick,circle, minimum size = 1.1cm, fill =DAI] at ({360/\n *3 +90}:\radius) (v5)  {\scriptsize DAI};
	\node[draw, thick,circle, minimum size = 1.1cm, fill =USDT] at ({360/\n *2+90}:\radius) (v1) {\scriptsize USDT};
	\node[draw, thick,circle, minimum size = 1.1cm, fill =USDC] at ({360/\n *1+90}:\radius) (v2) {\scriptsize USDC};
	\node[draw, thick,circle, minimum size = 1.1cm, fill =ETH] at ({360/\n *4+90}:\radius) (v3) {\scriptsize ETH};

    \path[<->, draw]
    (v1) edge[thick] node {}(v2)
    (v1) edge[thick, bend left = 10] node {}(v3)
    (v1) edge[dashed, thick, bend right = 5] node {}(v3)
    (v1) edge[thick] node {}(v5)
    
    (v2) edge[ thick, bend left = 10] node {}(v3)
    (v3) edge[dashed, thick, bend left = 5] node {}(v2)
    
    (v5) edge[dashed, thick, bend left = 10] node {}(v3)

    (v3) edge[thick, bend left = 10] node {}(v5)

    (v2) edge[thick] node {}(v5);

    
    \end{tikzpicture}
    \caption{Network during period (1)}\label{fig:arb_cyc_4_coins}
    \end{subfigure}
    \hfill
    \begin{subfigure}[b]{0.48\linewidth}
    \centering
    \begin{tikzpicture}[scale =0.65]

    \def \n {5}
	\def \radius {2.3cm}
	\def \margin {15} 

	\node[draw, thick,circle, minimum size = 1.1cm, fill =DAI] at ({360/\n *3+90}:\radius) (v5)  {\scriptsize DAI};
	\node[draw, thick,circle, minimum size = 1.1cm, fill =USDT] at ({360/\n *2+90}:\radius) (v1) {\scriptsize USDT};
	\node[draw, thick,circle, minimum size = 1.1cm, fill =USDC] at ({360/\n *1+90}:\radius) (v2) {\scriptsize USDC};
	\node[draw, thick,circle, minimum size = 1.1cm, fill =ETH] at ({360/\n *4+90}:\radius) (v3) {\scriptsize ETH};
	\node[draw, thick,circle, minimum size = 1.1cm, fill =BTC] at ({360/\n *5+90}:\radius) (v4) {\scriptsize BTC};
    
    
    \path[<->, draw]
    (v1) edge[ thick] node {}(v2)
    (v1) edge[ thick, bend left = 10] node {}(v3)
    (v1) edge[dashed, thick, bend right = 5] node {}(v3)
    (v1) edge[ thick] node {}(v5)

    (v3) edge[ thick, bend left = 10] node {}(v2)
    (v3) edge[dashed, thick, bend right = 5] node {}(v2)

    (v3) edge[thick, bend left = 10] node {}(v5)
    (v3) edge[dashed, thick, bend right = 5] node {}(v5)

    (v2) edge[ thick,] node {}(v4)
    (v3) edge[ thick, bend left = 10] node {}(v4)
    (v3) edge[dashed, thick, bend right = 10] node {}(v4)
    
    (v2) edge[thick] node {}(v5)

    ;
    
    \end{tikzpicture}
    \caption{Network during period (2)}\label{fig:arb_cyc_5_coins}
    \end{subfigure}
    \caption{Network of pools considered during the two analyzed time periods. Cryptocurrencies are nodes and edges represent pools between the respective cryptocurrencies. Uniswap pools are represented by solid lines and SushiSwap pools are represented by dashed lines.}
    \label{fig:arb_cyc_coins}
\end{figure}

\end{document}